\documentclass[twocolumn,showpacs,superscriptaddress,preprintnumbers,prd]{revtex4}
\usepackage{amsmath,amssymb}
\usepackage{graphicx}
\usepackage{dcolumn}
\usepackage{epsfig,color}
\usepackage{bm}
\usepackage{verbatim}

\begin{document}
\title{Open-flavor strong decays of open-charm and open-bottom mesons in the $^3P_0$ model}
\author{J. Ferretti}
\affiliation{Institute of Theoretical Physics, Chinese Academy of Sciences, Beijing 100190, China}
\affiliation{Dipartimento di Fisica and INFN, ``Sapienza" Universit\`a di Roma, Piazzale A. Moro 2, 00185 Roma, Italy}
\author{E. Santopinto}\email[]{santopinto@ge.infn.it}
\affiliation{INFN, Sezione di Genova, via Dodecaneso 33, 16146 Genova, Italy}

\begin{abstract}
We provide results for the open-flavor strong decays of open-charm ($D$ and $D_{\rm s}$) and open-bottom ($B$, $B_{\rm s}$ and $B_{\rm c}$) mesons. The decays are calculated in a modified version of the $^3P_0$ pair-creation model, assuming harmonic oscillator wave functions. The spectra of open-charm and open-bottom mesons used in the calculations are computed within Godfrey and Isgur's relativized quark model. Quantum number assignments are also provided. Our results are compared with the existing experimental data. 
\end{abstract}
\maketitle

\section{Introduction}
Since the discovery of the $J/\Psi$ and $\Upsilon$ resonances in the 1970s, heavy meson physics (including the physics of charmonia \cite{Choi:2003ue,Aaij:2013zoa}, bottomonia \cite{Aad:2011ih,Abazov:2012gh}, open-charm \cite{Aubert:2003fg,Besson:2003cp} and open-bottom \cite{Bebek:1980zd,LeeFranzini:1990gy,Abe:1998fb} mesons) has been been extensively studied, and still is subject of intensive theoretical and experimental research \cite{Nakamura:2010zzi,Brambilla:2010cs}. 
Recently, both the charmonium and bottomonium spectra have been enriched by the discovery of new particles \cite{Nakamura:2010zzi,Brambilla:2010cs}; also the knowledge of open-charm and open-bottom mesons has improved substantially with the experimental observation of new resonances, including the $D_0^*(2400)$ \cite{Link:2003bd,Abe:2003zm}, $D_1(2430)^0$ \cite{Abe:2003zm} and $B_1(5721)$ \cite{Abazov:2007vq,Aaltonen:2008aa}. See Table \ref{tab:total-DeB}.
The properties and quantum numbers of a large part of the newly-observed open-charm and open-bottom mesons are still not well established. Some examples are $D_J^*(2600)$ \cite{delAmoSanchez:2010vq,Nakamura:2010zzi}, $D(2740)^0$ \cite{Nakamura:2010zzi,Aaij:2013sza} and $B_J(5970)^0$ \cite{Nakamura:2010zzi,Aaltonen:2013atp}.
This has led to remarkable theoretical efforts to provide the experimentalists predictions for the spectrum, decay modes, and so on, and also attempts to make quark model assignments for new observed states. 

Important information on mesons can be extracted from their possible decay modes, including electromagnetic, weak and strong decays. The possibility to provide a theoretical description of (open- and hidden-flavor) strong decays relies mainly on phenomenological models, because the operators that describe the strong transitions between hadrons, arising from non-perturbative QCD, are essentially unknown. 
In the open-flavor case, they include, among other things, ``hadrodynamic" models, pair-creation models, elementary meson emission models, heavy meson effective theories and Lagrangian approaches with heavy-quark and chiral symmetries \cite{Kilian:1992hq,Colangelo:1995ph,Casalbuoni:1996pg,Colangelo:2007ds,Wang:2010ydc,Colangelo:2012xi,Wang:2013tka,Wang:2014cta,Wang:2016ewb}. For a review, see \cite{Capstick:2000qj}. 

In this paper, we focus on the $^3P_0$ pair-creation model, in which the decays proceed via the production of $q \bar q$ pairs with vacuum quantum numbers, i.e. $J^{PC} = 0^{++}$, somewhere in the hadronic medium \cite{Micu}. 
An important feature of the $^3P_0$ model, apart from its simplicity, is that it provides the gross features of several transitions with only one free parameter, the pair-creation strength $\gamma_0$, which is a free constant to be fitted to the experimental data. 
More recent studies have also discussed the possibility of substituting the constant pair-creation vertex of the model with a more refined one \cite{Kokoski:1985is,Kumano:1988ga,Stancu:1988gb,Geiger:1994,Kalashnikova:2005ui,Segovia:2012cd}.
Extensively applied to the study of open-flavor strong decays of light mesons \cite{Kokoski:1985is,Blundell:1995ev,Ackleh:1996yt} and baryons \cite{Capstick:1992th,Strong2015,LeYaouanc}, the $^3P_0$ pair-creation model has also been used to compute the decays of charmonia \cite{LeYaouanc02,Barnes:2005pb,Ferretti:2013faa,Ferretti:2014xqa}, bottomonia \cite{Ferretti:2013vua,Godfrey:2015dia}, open-charm \cite{Close:2005se,Segovia:2012cd,Godfrey:2015dva} and open-bottom \cite{Godfrey:2016nwn} mesons. 

The aim of the present paper is to provide a classification of open-charm and open-bottom mesons in terms of their masses, calculated within Godfrey and Isgur's relativized model \cite{Godfrey:1985xj,Godfrey:2004ya}, open-flavor amplitudes, evaluated within a modified version of the $^3P_0$ pair-creation model \cite{Ferretti:2013faa,Ferretti:2014xqa,Ferretti:2013vua,bottomonium}, and quantum number assignments, carried out by comparing our predictions to the existing data. 

As widely shown by previous quark model calculations, we expect to obtain a good overall description of the properties of these mesons, with the possible exception of states close to meson-meson decay thresholds, like $D_0^*(2400)$ and $D_{s0}^*(2317)$ \cite{Hwang:2004cd}.
Indeed, it is well known that the quenched approximation may fail for states in the region around the opening of meson-meson decay thresholds, where it is believed that continuum-coupling effects may play an important role \cite{Ferretti:2013faa,Ferretti:2014xqa,Guo:2011dv,Cardoso:2014xda,Guo:2016yxl,Lu:2016mbb,Ferretti:2018tco}.
A study of these particular states in the context of coupled-channel models will be addressed in a future publication.

\section{Formalism}
\subsection{$^3P_0$ pair-creation model}
\label{3P0-formulas}
In the $^3P_0$ pair-creation model, the open-flavor strong decay of a hadron $A$ into hadrons $B$ and $C$ takes place in its rest frame, via the creation of an additional $q \bar q$ pair characterized by $J^{PC} = 0^{++}$ quantum numbers \cite{Micu,LeYaouanc,Roberts:1992}. 
The decay widths $A \rightarrow BC$ are calculated as \cite{Micu,LeYaouanc,Ackleh:1996yt} 
\begin{equation}
	\label{eqn:3P0-decays-ABC}
	\Gamma_{A \rightarrow BC} = \Phi_{A \rightarrow BC}(q_0) \sum_{\ell} 
	\left| \left\langle BC q_0  \, \ell J \right| T^\dag \left| A \right\rangle \right|^2 \mbox{ }.
\end{equation}
The coefficient  
\begin{equation}
	\label{eqn:rel-PSF}
	\Phi_{A \rightarrow BC}(q_0) = 2 \pi q_0 \frac{E_b(q_0) E_c(q_0)}{M_a}  \mbox{ },
\end{equation}
depending on the relative momentum $q_0$ between $B$ and $C$ and the energies of the two decay products, $E_{b,c} = \sqrt{M_{b,c}^2 + q_0^2}$, is the phase space factor for the decay.
We assume harmonic oscillator wave functions, depending on a single oscillator parameter $\alpha$. 
The final state is characterized by the relative orbital angular momentum $\ell$ between $B$ and $C$ and a total angular momentum $\vec{J} = \vec{J}_b + \vec{J}_c + \vec{\ell}$.
%%%%%%%%%%%%%%%%%%%%%%%%%%%%%%%%%%%%%%%%%%%%%%%%%%%%%%%%%%%%%%%%%%%%%%%%%%
\begin{table}[htbp]  
\begin{center}
\begin{tabular}{cccc} 
\hline 
\hline
State                 & $J^P$     & Exp. Mass [MeV] & $\Gamma^{\rm exp}_{\rm tot}$ [MeV]  \\
\hline
$D^*(2007)^0$      & $1^-$    & $2006.85\pm0.05$ & $< 2.1$  \\
$D_0^*(2400)^0$       & $0^+$    & $2318\pm29$ & $267\pm40$ \\ 
$D_1(2420)^0$       & $1^+$    & $2420.8\pm0.5$ & $31.7\pm2.5$ \\
$D_1(2430)^0$ $\$$ & $1^+$    & $2427\pm26\pm25$ & $384^{+107}_{-75}\pm74$ \\
$D_{s1}(2460)^\pm$   & $1^+$    & $2459.5\pm0.6$ & $< 3.5$ \\
$D_{s1}(2536)^\pm$   & $1^+$    & $2535.10\pm0.06$ & $0.92\pm0.03\pm0.04$ \\
$D_2^*(2460)^0$  & $2^+$    & $2460.7\pm0.4$ & $47.5\pm1.1$  \\
$D_0(2550)^0$ $\$$ & $0^-$    & $2564\pm20$ & $135\pm17$  \\
$D_{s2}^*(2573)^\pm$    & $2^+$    & $2569.1\pm0.8$ & $16.9\pm0.8$   \\
$D_{s1}^*(2700)^\pm$  & $1^-$    & $2708.3^{+4.0}_{-3.4}$ & $120\pm11$ \\   
$D_{s1}^*(2860)^\pm$ $\$$ & $1^-$  & $2859\pm12\pm24$ & $159\pm23\pm77$ \\
$D_3(2750)$ $\$$ & $3^-$  & $2763.5\pm3.4$ & $66\pm5$ \\ 
$D_{s3}^*(2860)$ $\$$ & $3^-$ & $2860.5\pm2.6\pm6.5$ & $53\pm7\pm7$ \\ \hline
$B_1(5721)^0$ & $2^+$  &  $5726.0\pm1.3$ &  $27.5\pm3.4$ \\
$B_2^*(5747)^0$ & $2^+$  &  $5739.5\pm0.7$ &  $24.2\pm1.7$ \\
$B_{s1}(5830)^0$ & $1^+$ & $5828.63\pm0.27$ & $0.5\pm0.3\pm0.3$      \\  
$B_{s2}^*(5840)^0$ & $2^+$ & $5839.85\pm0.17$ & $1.47\pm0.33$      \\  
\hline 
\hline
\end{tabular}
\end{center}
\caption{Experimental total decay widths and masses of $D$, $D_s$, $B$ and $B_s$ mesons, extracted from the PDG \cite{Nakamura:2010zzi}. There is no data available for $B_{\rm c}$ states above the $BD$ threshold. States labeled by $\$$ are omitted from the PDG summary table \cite{Nakamura:2010zzi}.}
\label{tab:total-DeB}  
\end{table}
%%%%%%%%%%%%%%%%%%%%%%%%%%%%%%%%%%%%%%%%%%%%%%%%%%%%%%%%%%%%%%%%%%%%%%%%%%

Following Refs. \cite{Ferretti:2013faa,Ferretti:2014xqa,Ferretti:2013vua,bottomonium,self-energies}, we introduce a few changes in the $^3P_0$ model operator, $T^\dag$.
These modifications include the substitution of the pair-creation strength, $\gamma_0$, with an effective one \cite{Ferretti:2013faa,Ferretti:2014xqa,Ferretti:2013vua,bottomonium,Kalashnikova:2005ui,self-energies},
\begin{equation}
	\gamma_0^{\rm eff} = \frac{m_n}{m_i} \gamma_0  \mbox{ },
\end{equation}	
with $i = n$ (i.e. $u$ or $d$), $s$, $c$ and $b$ (see Table \ref{tab:3P0parameters}), to suppress heavy quark pair-creation.
Something similar was also done in Ref. \cite{Segovia:2012cd}, though the authors used a different form for $\gamma_0^{\rm eff}$.
We also introduce a Gaussian quark form-factor, with parameter $r_{\rm q}$, because the pair of created quarks has an effective size \cite{Ferretti:2013faa,Ferretti:2014xqa,Ferretti:2013vua,bottomonium,self-energies}.

The values of the pair-creation model parameters for the SU(4)$_{\rm f}$ and SU(5)$_{\rm f}$ sectors, reported in Table \ref{tab:3P0parameters}, are extracted from Refs. \cite{Ferretti:2013faa,Ferretti:2014xqa,Ferretti:2013vua}. These are the values that we use in our calculations.

%%%%%%%%%%%%%%%%%%%%%%%%%%%%%%%%%%%%%%%%%%%%%%%%%%%%%%%%%%%%%%%%%%%%%%%%%% 
\begin{table}[htbp]  
\begin{center}
\begin{tabular}{ccc} 
\hline 
\hline
Parameter  & Value in SU$_{\rm f}$(4) & Value in SU$_{\rm f}$(5)   \\ 
\hline
$\gamma_0$ & 0.510 & 0.732      \\  
$\alpha$   & 0.500 GeV & 0.500 GeV \\  
$r_{\rm q}$      & 0.335 fm & 0.335 fm  \\
$m_n$      & 0.330 GeV & 0.330 GeV \\
$m_s$      & 0.550 GeV & 0.550 GeV \\
$m_c$      & 1.50 GeV & 1.50 GeV  \\
$m_b$      & -- & 4.70 GeV   \\ 
\hline 
\hline
\end{tabular}
\end{center}
\caption{Pair-creation model parameters for SU$_{\rm f}$(4) and SU$_{\rm f}$(5) sectors, from Refs. \cite{Ferretti:2013faa,Ferretti:2014xqa,Ferretti:2013vua}.}
\label{tab:3P0parameters}  
\end{table}
%%%%%%%%%%%%%%%%%%%%%%%%%%%%%%%%%%%%%%%%%%%%%%%%%%%%%%%%%%%%%%%%%%%%%%%%%%

%%%%%%%%%%%%%%%%%%%%%%%%%%%%%%%%%%%%%%%%%%%%%%%%%%%%%%%%%%%%%%%%%%%
\begin{table*}
\begin{tabular}{cccccccccccccccc} 
\hline 
\hline 
State                 & $J^P$     & Mass [MeV] & $D \pi$     & $D^* \pi$ & $D \rho$ & $D^* \rho$ & $D \eta$ & $D^* \eta$ & $D \omega$ & $D^* \omega$ &  $D_s K$ & $D_s^* K$ & $D_s K^*$ & $D_s^* K^*$ \\ 
\hline
$D^*(2007)$ or $D_1(1^3S_1)$     & $1^-$    & 2038, $2007^\dag$ &  0      &   --          & --     &  --     &  --          &  --            & --                  &  --                  &      --        &   --           &   --           & --           \\  
$D_0^*(2400)$ or $D_0(1^3P_0)$     & $0^+$    & 2398, $2318^\dag$ &  66      &   --     & --      &  --      &  --          &  --      & --                  &  --                  &      --        &   --           &   --           & --           \\    
$D_1(2420)$ or $D_1(1P_1)$    & $1^+$    & 2456, $2421^\dag$ &  --           &   32        & --         &  --          &  --        &  --           & --                &  --               &     --        &   --         &   --        & --      \\ 
$D_1(2430)$ or $D_1(1P_1')$     & $1^+$    & 2467, $2427^\dag$ &  --           &   37         & --         &  --          &  --        &  --           & --                &  --               &     --        &   --         &   --        & --      \\         
$D_2^*(2460)$ or $D(1^3P_2)$     & $2^+$  & 2501, $2461^\dag$ &  6      &   2       & --           &  --             &  0          &  --            & --                  &  --        &      --        &   --           &   --           & --      \\ 
$D_0(2550)$ or $D(2^1S_0)$     & $0^-$    & 2582, $2564^\dag$ &  --            &   45          & --           &  --             &  --          &  0            & --                  &  --          &      --        &   --       &   --       & --       \\ 
$D_1(2^3S_1)$     & $1^-$    & 2645 &  18            &   36          & 0           &  --             &  6          &  5            & --                  &  --                  &      4        &   1           &   --           & --                     \\
$D_1(1^3D_1)$     & $1^-$    & 2816 &  20            &   13          & 13           &  1             &  10          &  5           & 4                  & 0                  &      6       &   2           &   --           & --                     \\ 
$D_2(1D_2)$     & $2^-$    & 2816 &  --            &   25      & 21        &  6        &  --          &  7      & 7        &  1           &      --       &   3   &   --          & --      \\
$D_3(2750)$ or $D_3(1^3D_3)$ & $3^-$ & 2833, $2764^\dag$ &  8            &   5          & 0          &  --             &  1          &  0            & 0     &  --                  &     0      &   0           &   --          & --               \\
$D_2(1D_2')$     & $2^-$    & 2845 &  --           &   26      & 23        &  5        &  --          &  8      & 8        &  1           &      --        &   4  &   --        & --              \\
$D_1(2P_1)$     & $1^+$    & 2924 &  --         &   26          & 20       &  33        &  --        &  7        & 7          &  11            &     --        &   4          &   6         & --         \\
$D_0(2^3P_0)$     & $0^+$    & 2931 &  18            &   --          & --           &  38             &  2          &  --            & --                  &  12                  &      0       &   --           &   --           & --                  \\ 
$D_2(2^3P_2)$     & $2^+$    & 2957 &  13            &   23          & 22           &  45             &  6          &  7            & 7                  &  16                &      4     &  4         &   1           & --    \\
$D_1(2P_1')$     & $1^+$    & 2961 &  --           &   21          & 14      &  29        &  --        &  5        & 5         &  9           &     --        &   3       &   8    & --             \\
$D_0(3^1S_0)$     & $0^-$    & 3067 &  --            &   1          & 4           &  38             &  --          &  1            & 1                  &  13       &      --        &   3     &   8           & 8                  \\
$D_1(3^3S_1)$     & $1^-$    & 3111 &  3            &   2          & 1           &  31             &  0          &  0            & 0                  &  11             &      0        &   1           &   5   & 15                     \\
$D_4(1^3F_4)$     & $4^+$    & 3113 &  11            &   8          & 4          &  36             &  2          &  1            & 1                  &  12                  &     1      &   0           &   0          & 1               \\
$D_2(1^3F_2)$     & $2^+$    & 3132 &  10            &   9          & 11          &  12             &  5          &  3            & 4                  &  4                  &      2       &   2           &   1           & 0                 \\
$D_2(2D_2)$     & $2^-$    & 3212 &  --         &   15    & 15       &  30         &  --        &  5            & 5           &  10          &   --        &   3        &   3      & 5      \\
$D_3(2^3D_3)$     & $3^-$    & 3226 &  8            &   14          & 16          &  21             &  4          &  5            & 5                  &  7                &     3      &   3           &   2         & 9              \\
$D_2(2D_2')$     & $2^-$    & 3248 &  --        &  14     & 13           &  32       &  --       &  4        & 4         &  11       &    --        &   3      &   3       & 4         \\
$D_1(3P_1)$     & $1^+$    & 3328 &  --           &   1      & 1     &  10       &  --        &  0        & 0         &  3         &     --        &   1      &   3       & 6           \\
$D_1(2^3D_1)$     & $1^-$    & 3231 &  7           &   2          & 0           &  51             &  1         &  0         & 0                  & 17                  &      0       &   0           &   1           & 4                     \\
$D_0(3^3P_0)$     & $0^+$    & 3343 &  1            &   --          & --           &  13            &  0          &  --            & --                  &  4                &      1       &   --           &   --           & 11                 \\
$D_2(3^3P_2)$     & $2^+$    & 3352 &  2            &   1     & 0          &  13           &  0      &  0         & 0                  &  5               &      0    &  1         &   1           &  6  \\
$D_1(3P_1')$     & $1^+$    & 3360 &  --           &   1    & 1       &  7      &  --        &  0     & 1        &  3        &     --        &   1       &   2       & 8           \\
$D_3(1^3G_3)$     & $3^-$    & 3398 &  5           &   2         & 7          &  15             &  2          &  2            & 2                  &  5                  &     1      &   1           &   1          & 1               \\
$D_0(4^1S_0)$     & $0^-$    & 3465 &  --            &   1          & 4           &  11             &  --          &  1            & 1                  &  4                  &      --        &   1           &   1           & 0                  \\
$D_4(2^3F_4)$     & $4^+$    & 3466 &  5            &   8          & 10          &  12             &  2          &  3            & 3                  &  4                  &     2      &   2        &   2         & 4              \\
$D_2(2^3F_2)$     & $2^+$    & 3490 &  3            &   1          & 0          &  38             &  1          &  0            & 0                 &  12                  &      0       &   0           &   0           & 5                 \\
$D_2(3D_2)$     & $2^-$    & 3566 &  --       &   1      & 1       &  3         &  --       &  0            & 0            &  1          &      --     &   1      &   1       & 3         \\
$D_3(3^3D_3)$     & $3^-$    & 3578 &  2            &   1          &  0          &  6             &  0          &  0            & 0                  &  2               &     0      &   0           &   1         & 2              \\
$D_2(3D_2')$     & $2^-$    & 3600 &  --      &   1     & 1        &  3         &  --     &  0         & 0          &  1      &      --        &   1     &   1         & 3          \\
\hline 
\hline
\end{tabular}
\caption{Open-flavor strong decay widths (in MeV) for $D$ states. Column 3 gives the values of the masses of the decaying mesons: when available, we use the experimental values from the PDG, denoted by the symbol $\dag$ \cite{Nakamura:2010zzi}; otherwise, we consider the predictions of the relativized QM for mesons \cite{Godfrey:1985xj}. Columns 4-15 show the decay width contributions (in MeV) from various channels, such as $D \pi$, $D^* \pi$, and so on. The values of the $^3P_0$ model parameters are given in Table \ref{tab:3P0parameters}. The symbol -- in the table means that a certain decay is forbidden by selection rules or that the decay cannot take place because it is below threshold. The calculated mixing angles are: $\theta_{\rm 1P} = 25.7^\circ$, $\theta_{\rm 2P} = 29.4^\circ$, $\theta_{\rm 3P} = 28.1^\circ$, $\theta_{\rm 1D} = 38.2^\circ$, $\theta_{\rm 2D} = 37.4^\circ$, $\theta_{\rm 3D} = 36.9^\circ$.}
\label{tab:strong-decays-D}  
\end{table*}
%%%%%%%%%%%%%%%%%%%%%%%%%%%%%%%%%%%%%%%%%%%%%%%%%%%%%%%%%%%%%%%%%%%%%%%%%%
\begin{table*}
\begin{tabular}{cccccccccccc} 
\hline 
\hline
State                 & $J^P$     & Mass [MeV] & $D K$     & $D^* K$ & $D K^*$ & $D^* K^*$ & $D_s \eta'$ & $D_s^* \eta'$ & $D_s \phi$ & $D_s^* \phi$  \\
\hline
$D_{\rm s1}(2460)$ or $D_{\rm s1}(1P_1)$     & $1^+$    & 2549, $2460^\dag$   &  --          &   -- (46)      & --       &  --       &  --       &  --      & --           &  --       \\
$D_{\rm s1}(2536)$ or $D_{\rm s1}(1P_1')$     & $1^+$   & 2556, $2535^\dag$ &  --          &   56 (54)     & --       &  --       &  --       &  --            & --   & --      \\          
$D_{\rm s2}^*(2573)$ or $D_{\rm s2}(1^3P_2)$  & $2^+$    & 2591, $2569^\dag$ &  4            &   0          & --           &  --             &  --          &  --            & --                  &  --            \\   
$D_{\rm s0}(2^1S_0)$     & $0^-$    & 2675 &  --            &   53          & --           &  --             &  --          &  --            & --                  &  --                  \\
$D_{\rm s1}^*(2700)$ or $D_{\rm s1}(2^3S_1)$     & $1^-$    & 2735, $2708^\dag$ &  28            &   42          & --           &  --             &  --          &  --            & --                  &  --            \\
$D_{\rm s1}^*(2860)$ or $D_{\rm s1}(1^3D_1)$     & $1^-$    & 2898, $2859^\dag$ &  43            &   23          & 13           &  --             &  --         &  --            & --               &  --          \\  
$D_{\rm s3}^*(2860)$ or $D_{\rm s3}(1^3D_3)$     & $3^-$    & 2916, $2861^\dag$ &  10 (14)          &  5 (8)        &  0 (1)           &  -- (5)             &  --          &  --            & --                 &  --            \\
$D_{\rm s2}(1D_2)$     & $2^-$    & 2900 &  --          &   40     & 31     &  1         &  --       &  --       & --              &  --      \\
$D_{\rm s2}(1D_2')$     & $2^-$    & 2926 &  --          &   43     & 34       &  1       &  --       &  --        & --             &  --      \\
$D_{\rm s0}(2^3P_0)$     & $0^+$    & 3005 &  28            &   --          & --           &  51     &  14          &  --            & --                  &  --            \\   
$D_{\rm s1}(2P_1)$     & $1^+$    & 3018 &  --          &   37       & 27         &  44           &  --       &  --      & 7            &  --        \\
$D_{\rm s1}(2P_1')$    & $1^+$    & 3038 &  --          &   31       & 20         &  35           &  --       &  --      & 10       &  --          \\
$D_{\rm s2}(2^3P_2)$     & $2^+$    & 3049 &  19          &   33        & 31           &  64             &  1          &  --            & 0                  &  --            \\
$D_{\rm s0}(3^1S_0)$     & $0^-$    & 3153 &  --            &   2          & 3           &  50             &  --          &  6            & 8                 &  3                 \\
$D_{\rm s4}(1^3F_4)$     & $4^+$    & 3190 &  17          &   11        &  5          &  60             &  0         &  0            & 0                 &  0          \\
$D_{\rm s1}(3^3S_1)$     & $1^-$    & 3194 &  6            &   3          & 0           &  39             &  2         &  5            & 5                &  11          \\
$D_{\rm s2}(1^3F_2)$     & $2^+$    & 3208 &  21          &   17        & 19           &  18             &  3          &  1            & 1                 &  4            \\
$D_{\rm s2}(2D_2)$     & $2^-$    & 3298 &  --          &   24     & 22     &  44    &  --       &  3        & 3            &  5        \\
$D_{\rm s1}(2^3D_1)$     & $1^-$    & 3306 &  13        &   3          & 1           &  74             &  1         &  2          &  1           &  4       \\
$D_{\rm s3}(2^3D_3)$     & $3^-$    & 3311 &  2          &   20        &  23        &  29             &  2          &  1         & 2             &  10            \\ 
$D_{\rm s2}(2D_2')$     & $2^-$    & 3323 &  --          &   22     & 21    &  47      &  --       &  3          & 3             &  5          \\
$D_{\rm s0}(3^3P_0)$     & $0^+$    & 3412 &  3            &   --          & --           &  15     &  2          &  --            & --                  &  12            \\
$D_{\rm s1}(3P_1)$     & $1^+$    & 3416 &  --          &   3   & 1     &  10     &  --       &  3         & 3           &  7           \\
$D_{\rm s1}(3P_1')$    & $1^+$    & 3433 &  --          &   2   & 1     &  7       &  --       &  3         & 2           &  8            \\
$D_{\rm s2}(3^3P_2)$     & $2^+$    & 3439 &  6          &   3        & 1           &  14             &  1          &  2            & 1                 &  6            \\
$D_{\rm s3}(1^3G_3)$     & $3^-$    & 3469 &  12          &   12        &  13          &  27             &  2          &  1            & 1                 &  1          \\
$D_{\rm s0}(4^1S_0)$     & $0^-$    & 3544 &  --            &   1         & 4           &  16             &  --          &  1            & 1                &  1                \\ 
$D_{\rm s4}(2^3F_4)$     & $4^+$    & 3544 &  7          &   12        &  15          &  18             &  2         &  1            & 2                 &  5          \\
$D_{\rm s2}(2^3F_2)$     & $2^+$    & 3562 &  8          &   3        &  2           &  59             &  0          &  0            & 0                 &  6            \\
$D_{\rm s2}(3D_2)$     & $2^-$    & 3650 &  --          &   3      & 2     &  3      &  --       &  1          & 1            &  3           \\
$D_{\rm s3}(3^3D_3)$     & $3^-$    & 3661 &  5          &   3        &  1          &  7             &  0          &  1            & 0                 &  3           \\
$D_{\rm s2}(3D_2')$     & $2^-$    & 3672 &  --          &   3   & 1      &  3       &  --       &  1       & 1          &  3          \\
\hline 
\hline
\end{tabular}
\caption{As Table \ref{tab:strong-decays-D}, but for $D_{\rm s}$ mesons. The calculated mixing angles are: $\theta_{\rm 1P} = 37.5^\circ$, $\theta_{\rm 2P} = 30.4^\circ$, $\theta_{\rm 3P} = 27.7^\circ$, $\theta_{\rm 1D} = 38.5^\circ$, $\theta_{\rm 2D} = 37.7^\circ$, $\theta_{\rm 3D} = 37.2^\circ$. In the $1P_1-1P_1'$ and $D_{\rm s3}^*(2860)$ cases, the values in parentheses are calculated by using the relativized QM predictions for the decaying meson masses.}
\label{tab:strong-decays-Ds}  
\end{table*}
%%%%%%%%%%%%%%%%%%%%%%%%%%%%%%%%%%%%%%%%%%%%%%%%%%%%%%%%%%%%%%%%%%%%%%%%%%

\subsection{Godfrey and Isgur's relativized quark model}
\label{Godfrey and Isgur's relativized constituent quark model}
The relativized quark model \cite{Godfrey:1985xj,Godfrey:2004ya} is based on an effective potential, whose dynamics is governed by a one-gluon exchange interaction at short distances plus a long-range linear confining one.

The Hamiltonian of the model is given by \cite{Godfrey:1985xj}
\begin{equation}
	H = \sqrt{q^2 + m_1^2} + \sqrt{q^2 + m_2^2} + V_{\rm conf} + V_{\rm hyp} + V_{\rm so}  \mbox{ },
	\label{eqn:H-GI}
\end{equation}
where $m_1$ and $m_2$ are the masses of the constituent quark and antiquark, $q$ their relative momentum (with conjugate coordinate $r$), $V_{\rm conf}$, $V_{\rm hyp}$ and $V_{\rm so}$ the confining, hyperfine and spin-orbit potentials, respectively.

The confining potential is the sum of three terms \cite{Godfrey:1985xj},
\begin{equation}
	V_{\rm conf} = - \left(\frac{3}{4} \mbox{ } c + \frac{3}{4} \mbox{ } br -\frac{\alpha_{\rm s}(r)}{r} \right) 
	\vec F_1 \cdot \vec F_2  \mbox{ },
\end{equation}
with $\left\langle q \bar q \right| \vec F_1 \cdot \vec F_2 \left| q \bar q \right\rangle = - \frac{4}{3}$.
The first term is a constant, the second a spin-independent linear confining one, with parameter $b$, and the third a Coulomb-like interaction.
The hyperfine interaction has the standard form \cite{Godfrey:1985xj}
\begin{equation}
	\label{eqn:Vhyp}
	\begin{array}{rcl}
	V_{\rm hyp} & = & -\frac{\alpha_{\rm s}(r)}{m_{1}m_{2}} \left[\frac{8\pi}{3} \vec S_{1} \cdot \vec S_{2} \mbox{ }
	\delta ^{3}(\vec r) \right. \\ 
	& + & \left. \frac{1}{r^{3}} \left( \frac{3 \mbox{ } \vec S_{1} \cdot \vec r \mbox{ } 
	\vec S_{2} \cdot \vec r}{r^{2}} - \vec S_{1} \cdot \vec S_{2}\right) \right] \mbox{ } 
	\vec F_{i} \cdot \vec F_{j}  \mbox{ }.
	\end{array}
\end{equation}
The spin-orbit potential \cite{Godfrey:1985xj}, 
\begin{equation}
	\label{eqn:Vso}
	V_{\rm so} = V_{\rm so,cm} + V_{\rm so,tp}  \mbox{ },
\end{equation}
is the sum of two contributions, where
\begin{subequations}
\begin{equation}
	\begin{array}{rcl}
	V_{\rm so,cm} & = & - \frac{\alpha_{\rm s}(r)}{r^{3}} \left( \frac{1}{m_{i}}+\frac{1}{m_{j}} \right) \\
	& & \left( \frac{\vec S_{i}}{m_{i}}+\frac{\vec S_{j}}{m_{j}} \right) \cdot \vec L 
	\;\;\vec F_{i}\cdot \vec F_{j}
	\end{array}
\end{equation}
is the color-magnetic term and
\begin{equation}
	V_{\rm so,tp} = - \frac{1}{2r}\frac{\partial V_{ij,{\rm conf}}}{\partial r} \left( \frac{\vec S_{i}} 
	{m_{i}^{2}}+\frac{\vec S_{j}}{m_{j}^{2}}\right) \cdot \vec L
\end{equation}
\end{subequations}
the Thomas-precession one. 

In the case of states characterized by quark and antiquark of unequal mass, charge conjugation is not a good quantum number. Therefore, states with different spins but the same angular momentum, $\left| n ^1L_J\right\rangle$ and $\left| n ^3L_J\right\rangle$, can mix via the spin-orbit interaction. For example, this happens in the case of $^1P_1$ and $^3P_1$ states, where we consider the linear combinations 
\begin{subequations}
\begin{equation}
	\begin{array}{l}
	\left| n P \right\rangle = \cos \theta_{nP} \left| n ^1P_1 \right\rangle + \sin \theta_{nP} \left| n ^3P_1 \right\rangle  \mbox{ }
	\end{array}
\end{equation}
and
\begin{equation}
	\begin{array}{l}
	\left| n P' \right\rangle = - \sin \theta_{nP} \left| n ^1P_1 \right\rangle + \cos \theta_{nP} \left| n ^3P_1 \right\rangle  \mbox{ }.
	\end{array}
\end{equation}
\end{subequations}
For more details, see Refs. \cite{Godfrey:1985xj,Godfrey:2004ya}. 

The spectrum of open-charm and open-bottom states, obtained by solving the eigenvalue problem of Eq. (\ref{eqn:H-GI}) with the values of the model paramaters of Ref. \cite{Godfrey:1985xj}, is reported in Tables \ref{tab:strong-decays-D}-\ref{tab:strong-decays-Bc}, third column. 

%%%%%%%%%%%%%%%%%%%%%%%%%%%%%%%%%%%%%%%%%%%%%%%%%%%%%%%%%%%%%%%%%%%%%%%%%%
\begin{table*}
\begin{tabular}{cccc} 
\hline 
\hline
Ratio                                                                                                                                    & $^3P_0$      & HMET and HQCS & Exp. \\
\hline
$\frac{\Gamma_{D_2^*(2460) \rightarrow D \pi}}{\Gamma_{D_2^*(2460) \rightarrow D^* \pi}}$ & 3.0 & 2.29 \cite{Wang:2016ewb}, 2.27 \cite{Colangelo:2012xi} & $1.5-3.0$ \cite{Albrecht:1989pa,Avery:1994yc,Chekanov:2008ac,delAmoSanchez:2010vq}  \\
$\frac{\Gamma_{D(2550)}^{\rm of}}{\Gamma_{D(2600)}^{\rm of}}$ & 0.64 & 0.85 \cite{Wang:2010ydc} & --  \\
$\frac{\Gamma_{D_{\rm s2}^*(2573) \rightarrow D^*K}}{\Gamma_{D_{\rm s2}^*(2573) \rightarrow DK}}$ & $\approx 0$ & 0.086 \cite{Colangelo:2012xi} & $< 0.33$ \cite{Nakamura:2010zzi} \\
$\frac{\Gamma_{D_{\rm s1}^*(2700) \rightarrow D^*K}}{\Gamma_{D_{\rm s1}^*(2700) \rightarrow DK}}$ & 1.5 & 0.91\cite{Colangelo:2012xi} & $0.91\pm0.13\pm0.12$ \cite{Nakamura:2010zzi} \\
$\frac{\Gamma_{D_{\rm s3}^*(2860) \rightarrow D^*K}}{\Gamma_{D_{\rm s3}^*(2860) \rightarrow DK}}$ & 0.50 $(0.57)^\dag$ & 0.39 \cite{Colangelo:2012xi} & $1.10\pm0.15\pm0.19$ \cite{Aubert:2009ah} \\
$\frac{\Gamma(B_2^*(5747) \rightarrow B^* \pi)}{\Gamma(B_2^*(5747) \rightarrow B \pi)}$ & 0.75 & 0.87 \cite{Colangelo:2012xi} & $1.0\pm 0.5 \pm 0.8$ \cite{Nakamura:2010zzi,Aaij:2015qla}  \\
$\frac{\Gamma(B_{s2}^*(5840) \rightarrow B^* K)}{\Gamma(B_2^*(5747) \rightarrow B K)}$ & $\approx 0$ & 0.07 \cite{Colangelo:2012xi} & $0.093\pm 0.013 \pm 0.012$ \cite{Nakamura:2010zzi}  \\
$\frac{\Gamma(B_3(1^3D_3) \rightarrow B^* \pi)}{\Gamma(B_3(1^3D_3) \rightarrow B \pi)}$ & 1.0 & 0.92 \cite{Colangelo:2012xi} & --  \\
$\frac{\Gamma(B_{s3}(1^3D_3) \rightarrow B^* K)}{\Gamma(B_{s3}(1^3D_3) \rightarrow B K)}$ & 0.94 & 0.815 \cite{Colangelo:2012xi} & --  \\
\hline 
\hline
\end{tabular}
\caption{Strong decay amplitude ratios. Our $^3P_0$ model results are compared to those of Heavy Meson Effective Theories (HMET) of Refs. \cite{Wang:2016ewb,Wang:2010ydc}, Lagrangian approaches with heavy quark and chiral symmetries (HQCS) of Refs. \cite{Colangelo:2012xi}, and the experimental data \cite{Nakamura:2010zzi,Albrecht:1989pa,Avery:1994yc,Chekanov:2008ac,delAmoSanchez:2010vq,Aaij:2015qla}. 
\\ $\dag$ See the caption of Table \ref{tab:strong-decays-Ds}.}
\label{tab:ratios}  
\end{table*}

\begin{table*}
\begin{tabular}{cccccccccccccccc} 
\hline 
\hline
State                 & $J^P$  & Mass [MeV]   & $B \pi$     & $B^* \pi$ & $B \rho$ & $B^* \rho$ & $B \eta$ & $B^* \eta$ & $B \omega$  & $B^* \omega$  
                          &   $B_s K$ & $B_s^* K$ & $B_s K^*$ & $B_s^* K^*$ \\
\hline
$B_1(5721)$ or $B_1(1P_1)$ & $1^+$   & 5777, $5726^\dag$ & --   &  $55^{\$}$      & --        &  --        &  --          &  --       &     --        &   --           &   --           & --         &   --           & --           \\
$B_2^*(5747)$ or $B_2(1^3P_2)$ & $2^+$  &  5796, $5740^\dag$ & 4      &   3      & --        &  --        &  --          &  --       &     --        &   --           &   --           & --         &   --           & --           \\ 
$B_0(1^3P_0)$     & $0^+$  &  5756 & 117           &   --          & --           &  --             &  --          &  --                            & 
                             --        &   --           &   --           & --                          &   --           & --                   \\
$B_1(1P_1')$  & $1^+$  & 5784 & --   &  73      & --        &  --        &  --          &  --       &     --        &   --           &   --           & --         &   --           & --           \\                             
$B_0(2^1S_0)$     & $0^-$  &  5905 & --           &   87          & --           &  --             &  --          &  4                            & 
                             --        &   --           &   --           & --                          &   --           & --                   \\
$B_1(2^3S_1)$     & $1^-$  &  5934 & 30           &   60          & --           &  --             &  7          &  6                            & 
                             --        &   --           &   3           & 1                          &   --           & --                   \\                             
$B_2(1D_2)$     & $2^-$   & 6095 & --   &  49      & 16      &  1        &  --          &  15      &     4       &   0         &   --           & 7        &   --           & --           \\                                                        
$B_3(1^3D_3)$     & $3^-$  &  6105 & 16           &   16          & 0           &  1             &  1          &  1                           & 
                             0        &   --           &  0           & 0                          &   --           & --                   \\
$B_1(1^3D_1)$     & $1^-$  &  6110 & 42           &  23          & 11          &  1             &  21         &  10                            & 
                             3        &   0           &   11           & 5                          &   --           & --                   \\ 
$B_2(1D_2')$     & $2^-$  & 6124 & --   &  50      & 17      &  2        &  --          &  16      &     5       &   0          &   --           & 8        &   --           & --           \\                             
$B_1(2P_1)$  & $1^+$  & 6197 & --   &  46      & 31      &  58       &  --          &  12      &     10       &   19          &   --           & 7       &   --           & --           \\                                                         
$B_2(2^3P_2)$     & $2^+$  &  6213 & 22           &   35          & 21          &  99             &  9          &  13                            & 
                             7        &   35           &   6           & 7                          &   --           & --                   \\ 
$B_0(2^3P_0)$     & $0^+$  &  6214 & 35           &   --          & --           &  70             &  3          &  --                            & 
                             --        &   22          &   0          & --                          &   --           & --                   \\ 
$B_1(2P_1')$  & $1^+$ & 6228 & --   &  42      & 29      &  48       &  --          &  10      &     10     &   15          &   --           & 5       &   --           & --           \\                              
$B_0(3^1S_0)$     & $0^-$  &  6334 & --           &   11          & 16           &  59            &  --          &  0                            & 
                             6        &  21           &   --           & 2                          &   11          & 6              \\ 
$B_1(3^3S_1)$     & $1^-$  &  6355 & 6           &   9          & 7           &  53            &  0          &  0                            & 
                             3       &   19          &   0           & 1                          &   9           & 16                  \\ 
$B_4(1^3F_4)$     & $4^+$  &  6364 & 19           &   20          & 4           &  80             &  3          &  3                            & 
                             1        &   26           &   1           & 1                          &   0           & 1                   \\ 
$B_2(1^3F_2)$     & $2^+$  &  6387 & 20           &   15          & 23          &  27            &  10         &  7                           & 
                             8        &   9           &   5           & 4                         &   1           & 0                 \\                              
$B_2(2D_2)$     & $2^-$  & 6450 & --   &  25      & 24      &  56       &  --          &  9      &     8       &   19          &   --           & 5        &   5         & 9          \\                                                          
$B_3(2^3D_3)$     & $3^-$  &  6459 & 12           &   18          & 26           &  37             &  6          &  9                           & 
                             8       &   12          &  4           & 5                          &   2           & 17                   \\  
$B_1(2^3D_1)$     & $1^-$  &  6475 & 15           &  6          & 0          &  100             &  3         &  1                            & 0 &
                             33        &   0           &   0          & 2                          &   7                              \\
$B_2(2D_2')$    & $2^-$  & 6486 & --   &  25      & 22      &  60       &  --          &  8      &     7       &   20          &   --           & 4        &   6         & 8          \\                             
$B_1(3P_1)$  & $1^+$    & 6557 & --   &  7      & 3      &  9       &  --          &  0      &     1       &   3          &   --           & 1        &   5         & 11          \\                                                        
$B_2(3^3P_2)$     & $2^+$  &  6570 & 6           &   7          & 0          &  16             &  1          &  0                            & 
                             0        &   6           &   0           & 0                          &   3           & 10                  \\
$B_1(3P_1')$  & $1^+$   & 6585 & --   &  7      & 3      &  5       &  --          &  0      &     1       &   2          &   --           & 1        &   4         & 12          \\                             
$B_0(3^3P_0)$     & $0^+$  &  6590 & 6           &   --          & --           &  5             &  0          &  --                            & 
                             --        &   2          &   0          & --                          &   --           & 18                   \\ 
$B_3(1^3G_3)$     & $3^-$  &  6622 & 10           &   9          & 15           &  38             &  5          &  4                           & 
                             5       &   12          &  2           & 2                          &   2           & 2                   \\   
$B_4(2^3F_4)$     & $4^+$  &  6679 & 5           &   9          & 17           &  22             &  3          &  5                            & 
                             6        &   7          &   2           & 3                          &   2           & 7                   \\
$B_0(4^1S_0)$     & $0^-$  &  6687 & --           &   0          & 6           &  16            &  --          &  1                            & 
                             2        &  5          &   --           & 1                          &   2          & 2              \\
$B_2(2^3F_2)$     & $2^+$  &  6704 & 9           &   5          & 12          &  69            &  2         &  1                           & 
                             0        &   23           &   0           & 0                         &   0           & 9                 \\                              
$B_2(3D_2)$    & $2^-$  & 6767 & --   &  6      & 1      &  2       &  --          &  1      &     0       &   1          &   --           & 0        &   1         & 4          \\                                                         
$B_3(3^3D_3)$     & $3^-$  &  6775 & 5           &   5          & 1           &  6             &  1          &  1                           & 
                             0      &   2          &  0           & 0                          &   1           & 4                \\ 
$B_2(3D_2')$   & $2^-$  & 6800 & --   &  5      & 1      &  2       &  --          &  1      &     0       &   1          &   --           & 0        &   1         & 4          \\                             
$B_3(2^3G_3)$     & $3^-$  &  6909 & 5           &   4          & 2           &  38             &  1         &  1                           & 
                             0       &   13          &  0           & 0                         &   0           & 7                 \\ 
$B_0(4^3P_0)$     & $0^+$  &  6954 & 1           &   --          & --           &  3             &  0          &  --                            & 
                             --        &   1          &   0          & --                          &   --           & 3                   \\ 
$B_4(3^3F_4)$     & $4^+$  &  6966 & 3           &   4          & 1           &  3             &  1          &  1                            & 
                             0       &   1         &   0          & 0                          &   0           & 2                   \\
$B_4(4^3F_4)$     & $4^+$  &  7230 & 0           &   0          &  0           &  2             & 0         &  0                            & 
                             0        &   1          &   0           & 0                          &   0           & 0                  \\ 
\hline 
\hline
\end{tabular}
\caption{As Table \ref{tab:strong-decays-D}, but for $B$ mesons. The calculated mixing angles are: $\theta_{\rm 1P} = 30.3^\circ$, $\theta_{\rm 2P} = 32.3^\circ$, $\theta_{\rm 3P} = 31.6^\circ$, $\theta_{\rm 1D} = 39.7^\circ$, $\theta_{\rm 2D} = 39.0^\circ$, $\theta_{\rm 3D} = 38.6^\circ$. The width denoted by the symbol $\$$ is calculated by using the mass of the decaying meson from Godfrey-Isgur model. If we use the experimental value of the mass and do not introduce a mixing with $B_1(1P_1')$, we get a $B^* \pi$ width of 42 MeV. In this latter case, the $B_1(1P_1)-B_1(1P_1')$ mixing is not introduced because the masses of the two mesons differ by 60 MeV, approximately.}
\label{tab:strong-decays-B}  
\end{table*}
%%%%%%%%%%%%%%%%%%%%%%%%%%%%%%%%%%%%%%%%%%%%%%%%%%%%%%%%%%%%%%%%%%%%%%%%%%

\section{Results and discussion}
Below we discuss our results of Tables \ref{tab:strong-decays-D}--\ref{tab:strong-decays-Bc} for the open-flavor strong decays of open-charm ($D$ and $D_{\rm s}$) and open-bottom ($B$, $B_{\rm s}$ and $B_{\rm c}$) mesons. 
The decays are computed in the $^3P_0$ pair-creation model formalism of Sec. \ref{3P0-formulas}, with the values of the model parameters of Table \ref{tab:3P0parameters} and Refs. \cite{Ferretti:2013faa,Ferretti:2014xqa,Ferretti:2013vua}. There, the parameters of the SU$_{\rm f}$(4) and SU$_{\rm f}$(5) sectors were fitted to the charmonium and bottomonium strong decay amplitudes, respectively.
When available, we calculate the amplitudes by using the experimental values of the meson masses, extracted from the PDG \cite{Nakamura:2010zzi}; otherwise, we use the relativized QM predictions reported in the third column of Tables \ref{tab:strong-decays-D} and \ref{tab:strong-decays-Ds}, \ref{tab:strong-decays-B}--\ref{tab:strong-decays-Bc}. 
See also Table \ref{tab:total-DeB}, which shows the existing experimental data for the total widths of $D$, $D_{\rm s}$, $B$ and $B_{\rm s}$ resonances. There is no data available for higher $B_{\rm c}$ resonances \cite{Nakamura:2010zzi}.
Our theoretical results reproduce the global trend of the PDG data \cite{Nakamura:2010zzi} with a few exceptions. 

In more detail, starting from the $D$ sector, our result for the open-flavor width of the $D^*(2007)^0$, $\Gamma^{\rm th}_{\rm of} = 4$ keV, is compatible with the total experimental width $\Gamma^{\rm exp}_{\rm tot} < 2.1$ MeV \cite{Nakamura:2010zzi}. A more refined prediction would require the introduction of coupled-channel effects, the mass of $D^*(2007)^0$ being very close to $D \pi$ threshold. The same applies to $D_0^*(2400)$, where the presence of higher Fock components in the meson wave function may lower the relativized QM prediction for the mass, 2398 MeV, down to the experimental value, $2318\pm29$ MeV, and also contribute to the open-flavor amplitude.
In the $D_1(2420)$ case, which should mainly decay into $D^* \pi$ with the possible chain $D^* \pi \rightarrow D \pi \pi$, our $^3P_0$ model prediction is compatible with the data, while this is not true for $D_1(2430)$, being $\Gamma^{\rm th}_{\rm of} \ll \Gamma^{\rm exp}_{\rm tot}$. Nevertheless, it is worth noting that, in this second case, the experimental error is still very large; moreover, if $D_1(2420)$ and $D_1(2430)$ are mixed by spin-orbit forces, their open-flavor widths are likely to be of the same order of magnitude. 
Our results for the total open-flavor widths of $D_2^*(2460)$ and $D_0(2550)$ are compatible with the present experimental data, being $\Gamma^{\rm th}_{\rm of} < \Gamma^{\rm exp}_{\rm tot}$. 
In the $D_2^*(2460)$ case, it is also interesting to calculate the ratio between the kinematically allowed open-flavor decays, namely ${\mathcal R}_{D_2^*(2460)}^{\rm of} = \Gamma_{D_2^*(2460) \rightarrow D \pi} / \Gamma_{D_2^*(2460) \rightarrow D^* \pi}$.
Our estimate, ${\mathcal R}_{D_2^*(2460)}^{\rm of} \approx 3$, is close to the heavy meson effective theory one of Ref. \cite{Wang:2016ewb}, i.e. 2.29. 
Our result can also be compared to the existing experimental data: $3.0\pm1.1\pm1.5$ \cite{Albrecht:1989pa}, $2.2\pm0.7\pm0.6$ \cite{Avery:1994yc}, $2.8\pm0.8^{+0.5}_{-0.6}$ \cite{Chekanov:2008ac}, $1.47\pm0.03\pm0.16$ \cite{delAmoSanchez:2010vq}.
Another interesting information is the ratio between the total decay widths of $D(2550)$ and $D(2600)$, under the hypothesis that $D(2600)$ has $2^3S_1$ quantum numbers.
Also in this second case our result, ${\mathcal R}_{D(2550)/D(2600)}^{\rm of} = \Gamma_{D(2550)}^{\rm of} / \Gamma_{D(2600)}^{\rm of} = 0.64$, is similar to the heavy meson effective theory one of Ref. \cite{Wang:2010ydc}, ${\mathcal R}_{D(2550)/D(2600)}^{\rm of} = 0.85$.

Moving to the $D_{\rm s}$ sector, our predictions for $D_{\rm s1}(2460)$ are compatible with the data \cite{Nakamura:2010zzi}, while those for $D_{\rm s1}(2536)$ are not. 
The former meson has a narrow width and mainly decays to $D_{\rm s}^*$ via photon or $\pi^0$ emission, which are normally suppressed decay modes \cite{Besson:2003cp}.
Because of the large mass difference between $D_{\rm s1}(2460)$ and $D_{\rm s1}(2536)$, which cannot be explained in terms of hyperfine or spin-orbit splittings, these mesons may have exotic nature. 
Our results for the total widths of $D_{\rm s2}^*(2573)$, $D_{\rm s1}^*(2700)$, $D_{\rm s1}^*(2860)$ and $D_{\rm s3}^*(2860)$ are compatible with the experimental data \cite{Nakamura:2010zzi}, being $\Gamma^{\rm th}_{\rm of} < \Gamma^{\rm exp}_{\rm tot}$. 
We cannot say much on the single channels, as the PDG only provides some preliminary results for a few branching fractions, except that, in the $D_{\rm s2}^*(2573)$ case, our predictions are compatible with $\Gamma(D^*K)/\Gamma(DK) < 0.33$ \cite{Nakamura:2010zzi}. 
In the $D_{\rm s3}^*(2860)$ case, we also show predictions extracted by using the relativized QM mass for the decaying meson because: I) There is a large difference between experimental and calculated masses; II) The experimental data are not very reliable as, at the moment, the state is excluded from the PDG summary table \cite{Nakamura:2010zzi}.

Finally, we discuss our predictions for the $B$ and $B_{\rm s}$ sectors. 
Our results for the total open-flavor widths of $B_2^*(5747)$ and $B_{s2}^*(5840)$ and for the ratio $\frac{\Gamma(B_2^*(5747) \rightarrow B^* \pi)}{\Gamma(B_2^*(5747) \rightarrow B \pi)}$ are compatible with the experimental data \cite{Nakamura:2010zzi,Aaij:2015qla}. 
By contrast, our result for the open-flavor width of $B_{s1}(5830)$ is incompatible with the data. 
Our prediction is very sensitive to the value of the decaying meson mass -- as $B_{s1}(5830)$ is close to the $B^* K$ threshold -- and thus a few MeV mass difference can produce large deviations in the calculated decay amplitude.
It is also interesting to discuss the possible quantum number assignments for $B_J(5970)^0$, whose width is $81\pm12$ MeV \cite{Nakamura:2010zzi}.
Following Ref. \cite{Wang:2014cta}, the three possible assignments are $2^3S_1$, $1^3D_3$ and $1^3D_1$. 
According to our results, the most likely is $2^3S_1$.

%%%%%%%%%%%%%%%%%%%%%%%%%%%%%%%%%%%%%%%%%%%%%%%%%%%%%%%%%%%%%%%%%%%%%%%%%% 
\begin{table*}
\begin{tabular}{ccccccccccc} 
\hline 
\hline
State                 & $J^P$   & Mass [MeV] & $B K$ & $B^* K$ & $B K^*$ & $B^* K^*$ & $B_s \eta'$ & $B_s^* \eta'$ & $B_s \phi$ & $B_s^* \phi$ \\
\hline
$B_{s1}(5830)$ or $B_{s1}(1P_1)$  & $1^+$  & 5857, $5829$ &  --   &  85 (30) &  --      &  --      &  --    &  --   &  --   &  --    \\
$B_{s0}(1^3P_0)$  & $0^+$  & 5830 & 208    &   --      & --          &  --           &  --                      &  --                & --                 &  --    \\ 
$B_{s2}^*(5840)$ or $B_{s2}(1^3P_2)$  & $2^+$ & 5875, $5840^\dag$ &   1      &   0                        & --                 &  --  &  --             &  --                & --                 &  --     \\
$B_{s1}(1P_1')$  & $1^+$ & 5861 &  --   &  98 &  --      &  --      &  --    &  --   &  --   &  --    \\
$B_{s0}(2^1S_0)$  & $0^-$ & 5985 &  --      &   106      & --          &  --           &  --             &  --                & --                 &  --     \\ 
$B_{s1}(2^3S_1)$  & $1^-$ & 6013 &   46   &   81        & --          &  --             &  --            &  --                 & --                 &  --    \\ 
$B_{s2}(1D_2)$   & $2^-$   & 6169 &  --   &  82 &  3     &  --      &  --    &  --      &  --   &  --    \\
$B_{s3}(1^3D_3)$ & $3^-$ & 6178 &   17     &   16       &  0                         &  --   &  --             &  --                & --                 &  --      \\
$B_{s1}(1^3D_1)$ & $1^-$  & 6181 &   89     &   47       &  1           &  --            &  --            &  --                 & --                 &  --   \\
$B_{s2}(1D_2')$  & $2^-$   & 6196 &  --   &  86 &  3     &  --      &  --    &  --      &  --   &  --    \\ 
$B_{s0}(2^3P_0)$ & $0^+$ & 6279 &  52      &   --      & --          &  71            &  --             &  --                & --                 &  --     \\ 
$B_{s1}(2P_1)$  & $1^+$   & 6279 &  --   &  70 &  47   &  81   &  --    &  --      &  --   &  --    \\
$B_{s2}(2^3P_2)$  & $2^+$ & 6295 &  35     &   56       &  24         &  172            &  --             &  --                & --                 &  --    \\ 
$B_{s1}(2P_1')$  & $1^+$  & 6296 &  --   &  63 &  49   &  51   &  --    &  --      &  --   &  --    \\
$B_{s0}(3^1S_0)$  & $0^-$  & 6409 &  --      &   16          & 24          &  92           &  --             &  6                &  4                 &  --      \\  
$B_{s1}(3^3S_1)$  & $1^-$ & 6429 &   10     &   21       &  10         &  84           & 5            &  6                 & 6                 &  --    \\ 
$B_{s4}(1^3F_4)$  & $4^+$ & 6431 &   28     &   27       &   4          &   125             &  0             &  0                & 0                 &  --           \\ 
$B_{s2}(1^3F_2)$  & $2^+$ & 6453 &   46     &   34       &  39         &   38          &  3             &  1                & 1                 &  0  \\ 
$B_{s2}(2D_2)$   & $2^-$    & 6526 &  --   &  44 &  36     &  83      &  --    &  6      &  7   &  9    \\
$B_{s3}(2^3D_3)$ & $3^-$ & 6534 &   18     &   29       &  39         &  53          &    2             &  1                & 1                 &  20      \\ 
$B_{s1}(2^3D_1)$ & $1^-$ & 6542 &   31     &   13       &  0           &  145        &  4            &  3                 & 4                 &  6   \\ 
$B_{s2}(2D_2')$  & $2^-$    & 6553 &  --   &  43 &  33     &  87      &  --    &  7      &  7   &  8    \\
$B_{s1}(3P_1)$  & $1^+$   & 6635 &  --   &  16 &  3     &  10      &  --    &  5      &  5   &  12    \\
$B_{s0}(3^3P_0)$ & $0^+$ & 6638 &  13      &   --      & --          &  10           &  5               &  --                & --                 &  21     \\ 
$B_{s2}(3^3P_2)$ & $2^+$ & 6647 &  12     &   14       &  0           &  19            &  2             &  4                &  4                 &  11     \\ 
$B_{s1}(3P_1')$  & $1^+$  & 6650 &  --   &  14 &  4     &  5        &  --    &  5      &  5   &  14    \\
$B_{s3}(1^3G_3)$ & $3^-$ & 6685 &   26     &   22       &  31         &   67           &  4              &  2                & 2                 &  2  \\ 
$B_{s4}(2^3F_4)$  & $4^+$ & 6747 &  7      &   13       &  26         &    37          &  3             &  3                & 3                 &  9            \\
$B_{s0}(4^1S_0)$  & $0^-$ & 6757 &  --      &   2          & 6            &  22          &  --             &  2                 &  2                  &  2      \\  
$B_{s2}(2^3F_2)$  & $2^+$ & 6768 &   21     &   13       &   4          &  110          &   0             &  1               & 1                 &  11  \\ 
$B_{s2}(3D_2)$     & $2^-$  & 6841 &  --   &  15 &  4     &  5      &  --    &  2      &  2   &  5    \\
$B_{s3}(3^3D_3)$ & $3^-$ & 6848 &  10     &   12       &  4           &  7              &  0             &  1                & 1                 &  6      \\
$B_{s2}(3D_2')$    & $2^-$  & 6864 &  --   &  14 &  3     &  5      &  --    &  2      &  2   &  5    \\
$B_{s0}(4^3P_0)$ & $0^+$ & 6949 &   2       &   --      & --          &  10           &  1               &  --                & --                 &   4     \\  
$B_{s3}(2^3G_3)$ & $3^-$ & 6970  &   15     &   10       &  6           &   62          &  0             &  0                & 0                 &  10   \\ 
$B_{s4}(3^3F_4)$ & $4^+$ & 7034 &   8      &    9        &   5          &     5            &  0             &  0                & 0                 &  3            \\
$B_{s4}(4^3F_4)$  & $4^+$ & 7297 &   3      &    2        &   0          &     3            &  0             &  0                & 0                 &  1            \\
\hline 
\hline
\end{tabular}
\caption{As Table \ref{tab:strong-decays-D}, but for $B_s$ mesons. The calculated mixing angles are: $\theta_{1P} = 39.1^\circ$, $\theta_{2P} = 33.1^\circ$, $\theta_{3P} = 31.6^\circ$, $\theta_{1D} = 40.0^\circ$, $\theta_{2D} = 39.5^\circ$, $\theta_{3D} = 39.1^\circ$. In the $B_{s1}(5830)$ case, the value in brackets is calculated by using the experimental value for the mass, without mixing with $B_{s1}(1P_1')$.} 
\label{tab:strong-decays-Bs}  
\end{table*}
 
\begin{table*}
\begin{tabular}{ccccccccccc} 
\hline 
\hline
State                 & $J^P$ & Mass [MeV] & $B D$ & $B^* D$ & $B D^*$ & $B^* D^*$ & $B_s D_s$ & $B_s^* D_s$  & $B_s D^*_s$  & $B_s^* D_s^*$ \\
\hline
$B_{c2}(2^3P_2)$ & $2^+$ & 7164 &   2    &   --       & --          &  --             &  --            &  --                & --                   &  --      \\ 
$B_{c0}(3^1S_0)$  & $0^-$ & 7249 &  --   &  107      & --          &  --             &  --            &  --                & --                  &  --      \\  
$B_{c2}(1^3F_2)$  & $2^+$ & 7269 & 90   &   29       & --          &  --             &  --            &  --                & --                  &  --      \\ 
$B_{c4}(1^3F_4)$  & $4^+$ & 7271 &  3    &    1        & --          &  --             &  --            &  --                & --                  &  --      \\ 
$B_{c1}(3^3S_1)$ & $1^-$ & 7272 &  13   &  64        & --          &  --            &  --             &  --                & --                  &  --      \\ 
$B_{c1}(2^3D_1)$ & $1^-$ & 7365 &  3    &   2         &  37         &   27           &   10           &  --                & --                  &  --      \\ 
$B_{c3}(2^3D_3)$ & $3^-$ & 7379 & 46   &   52       & 15          &  184          &  0              &  --                & --                  &  --      \\
$B_{c0}(3^3P_0)$  & $0^+$ & 7454 & 0    &   --       & --          &  157           &   4             &  --                & --                  &  --      \\
$B_{c3}(1^3G_3)$ & $3^-$ & 7474 & 93   &   66       & 45          &   27           &  5              &   1                 & --                  &  --      \\ 
$B_{c2}(3^3P_2)$  & $2^+$ & 7487 & 10   &    4        &  7           &  105          &   4             &  7                  &  0                  &  --      \\ 
$B_{c2}(2^3F_2)$  & $2^+$ & 7565 & 36   &   13       &  0           &  133          &   1             &  3                  & 4                   &   1      \\ 
$B_{c0}(4^1S_0)$ & $0^-$ & 7567 &  --   &   2         &  23         &  55            &  --            &  2                  & 2                    &  14      \\
$B_{c4}(2^3F_4)$ & $4^+$ & 7568 & 21   &   34       & 39          &  52            &   4             &  3                  & 0                   &  6      \\
$B_{c3}(3^3D_3)$ & $3^-$  & 7669 & 20   &   20       &  6           &   19           &   0             &   1                 &  3                  &  5      \\ 
$B_{c0}(4^3P_0)$ & $0^+$ & 7740 &  4    &   --       & --          &  22             &   3             &  --                & --                  &   1      \\   
$B_{c3}(2^3G_3)$ & $3^-$ & 7743 & 42   &   25       & 12          &  104          &   0             &   0                 &  1                  &  9      \\ 
$B_{c4}(3^3F_4)$  & $4^+$ & 7834 & 15   &   19       & 15          &  12            &   1             &  0                  & 0                   &  6      \\
$B_{c4}(4^3F_4)$  & $4^+$ & 8077 & 10   &   12       & 7            &  6              &   0             &  0                  & 0                   &  3      \\
\hline 
\hline
\end{tabular}
\caption{As Table \ref{tab:strong-decays-D}, but for $B_c$ mesons.} 
\label{tab:strong-decays-Bc}  
\end{table*}
%%%%%%%%%%%%%%%%%%%%%%%%%%%%%%%%%%%%%%%%%%%%%%%%%%%%%%%%%%%%%%%%%%%%%

\section{Summary and conclusion}
\label{Summary}
We computed the open-flavor strong decays of open-charm and open-bottom mesons within a modified version of the $^3P_0$ pair-creation model \cite{Micu,LeYaouanc}.

In the $^3P_0$ model, the open-flavor decays take place in the rest frame of the initial state, via the production of a light $q \bar q$ pair (i.e. $q = u$, $d$ or $s$) with $^3P_0$ quantum numbers.
Heavy quark pair production is heavily suppressed, as required by the phenomenology, by substituting the pair-creation strength, $\gamma_0$, with an effective one, $\gamma_0^{\rm eff}$ \cite{Ferretti:2013faa,Ferretti:2014xqa,Ferretti:2013vua,bottomonium,Kalashnikova:2005ui,self-energies}.
Moreover, the non-point-like nature of the pair of produced quarks is taken into account by introducing a quark form-factor \cite{Ferretti:2013faa,Ferretti:2014xqa,Ferretti:2013vua,bottomonium,Geiger-Isgur,Geiger:1996re,Bijker:2009up,Santopinto:2010zza,Bijker:2012zza,self-energies} into the model transition operator. 
The values of the $^3P_0$ model parameters in the SU$_{\rm f}$(4) and SU$_{\rm f}$(5) sectors were extracted from our previous studies on $c \bar c$ \cite{Ferretti:2013faa,Ferretti:2014xqa} and $b \bar b$ \cite{Ferretti:2014xqa,Ferretti:2013vua} meson spectroscopy and decays, where they were fitted to the existing experimental data \cite{Nakamura:2010zzi}. 

The open-charm and open-bottom meson spectra we needed in our calculation were predicted within Godfrey and Isgur's relativized quark model \cite{Godfrey:1985xj}. This is one of the most powerful tools for the study of $q \bar q$ meson spectroscopy, and provides a description of the meson spectrum in the light, strange, $c \bar c$, ..., sectors with a universal set of parameters; moreover, 30 years since its formulation, it still gives a good overall description of the experimental data. 

As discussed in our previous papers \cite{Ferretti:2013faa,Ferretti:2014xqa,Ferretti:2013vua}, there may be substantial deviations between the experimental values of the masses and QM predictions \cite{Godfrey:1985xj} in the case of resonances lying close to meson-meson decay thresholds. In these cases, continuum coupling effects may be important and determine a downward energy shift for the bare meson masses, thus improving the fit to the data; coupled-channel effects may also contribute to the open-flavor amplitudes. 
Such resonances may have an exotic nature, such as tetraquarks, meson-meson molecules or $q \bar q$ mesons plus continuum components. For example, this may be the case of $D(1^3S_1)$, $D_0^*(2400)$ and $D_{s0}^*(2317)$ \cite{Hwang:2004cd}, where QM predictions are incompatible with the present experimental data \cite{Nakamura:2010zzi}.  
The possible interpretations for suspected exotic open-charm and open-bottom mesons will be discussed in a future paper. 

In conclusion, we think that our predictions can be useful to the experimentalists in their study of the properties of open-charm and open-bottom mesons and in the search for new resonances.

\begin{appendix}

\section*{Flavor couplings in the $^3P_0$ pair-creation model}
In the following, we show how to calculate the SU$_{\rm f}$(5) flavor couplings of the $^3P_0$ pair-creation model. The SU$_{\rm f}$(4) couplings can be computed analogously.

We consider the transition $A \rightarrow BC$, where $A$, $B$ and $C$ are quark-antiquark mesons.
The SU$_{\rm f}$(5) flavor couplings can be written as the scalar product between initial, $\left| A (q_1 \bar q_2) \Phi_0 (q_3 \bar q_4) \right\rangle$, and final states, $\left| B(q_1 \bar q_4) C (q_3 \bar q_2) \right\rangle$. 
Here, $\Phi_0$ is the SU$_{\rm f}$(5) flavor singlet,
\begin{equation}
	\begin{array}{rcl}
	\left| \Phi_0 \right\rangle & = & \frac{1}{\sqrt{n_{\rm f}}} \sum_{i=1}^{n_{\rm f}} q_3^i \bar q_4^i  \\
	& = & \frac{1}{\sqrt 5} \left( \left| u \bar u \right\rangle 
	+ \left| d \bar d \right\rangle + \left| s \bar s \right\rangle 
	+ \left| c \bar c \right\rangle + \left| b \bar b \right\rangle \right)  		
 	\end{array}  \mbox{ },
\end{equation}
and $n_{\rm f} = 5$ is the dimension of the SU$_{\rm f}$ flavor group.
In general, two different diagrams can contribute to the flavor matrix element $\left\langle BC | A \Phi_0 \right\rangle$: in the first one, the quark in $A$ ends up in $B$, while in the second one it ends up in $C$. In the majority of cases, one of these two diagrams vanishes; however, for some matrix elements, both must be taken into account \cite{Ferretti:2013faa,Ferretti:2014xqa,Ferretti:2013vua,bottomonium}.
Finally, the flavor matrix elements can be calculated as:
\begin{equation}
	\label{eqn:FME-calc}
	\begin{array}{l}
	\left[\left\langle q_1 \bar q_4 \right| \otimes \left\langle q_3 \bar q_2\right| \right]
	\left[\left|q_1 \bar q_2 \right\rangle \otimes 
	\frac{1}{\sqrt{n_{\rm f}}} \sum_{i=1}^{n_{\rm f}} \left| q_3^i \bar q_4^i \right\rangle \right]  \\
	\hspace{0.5cm} = \frac{1}{\sqrt{n_{\rm f}}} \sum_{i=1}^{n_{\rm f}} \left[ \left\langle q_1 \bar q_4 q_3 \bar q_2 |
	q_1 \bar q_2 q_3^i \bar q_4^i  \right\rangle  \right. \\
	\hspace{0.45cm} \left. + \left\langle q_3 \bar q_2 q_1 \bar q_4 |
	q_1 \bar q_2 q_3^i \bar q_4^i  \right\rangle \right]
	\end{array}  \mbox{ }.
\end{equation}
As an example, we calculate the $B^0 \rightarrow B^0 \pi^0$ flavor coupling.
The flavor matrix element can be written as
\begin{equation}
	\label{eqn:B0B0pi0}
	\begin{array}{l}
	\left\langle B^0 \pi^0 | B^0 \Phi_0 \right\rangle_{\rm flavor} \\
	 \hspace{0.5cm} = - \frac{1}{\sqrt 10} \left[ \left\langle d \bar b d \bar d | d \bar b d \bar d \right\rangle 
	+ \left\langle d \bar d d \bar b | d \bar b d \bar d \right\rangle \right. \\
	\hspace{0.9cm} \left. - \left\langle d \bar b u \bar u | d \bar b u \bar u \right\rangle + ... \right]  =  - \frac{1}{\sqrt{10}}
	\end{array}  \mbox{ }.
\end{equation}
The only surviving contribution in Eq. (\ref{eqn:B0B0pi0}) is $\left\langle d \bar d d \bar b | d \bar b d \bar d \right\rangle$; the others, like $\left\langle d \bar b d \bar d | d \bar b d \bar d \right\rangle$ or $\left\langle d \bar b u \bar u | d \bar b u \bar u \right\rangle$, are null [see Eq. (\ref{eqn:FME-calc})]. In conclusion, after dividing Eq. (\ref{eqn:B0B0pi0}) by the corresponding SU(2) (isospin) Clebsch-Gordan coefficient, 
we get  
\begin{equation}
	\label{eqn:BBpi}
	\begin{array}{l}
		\left\langle B \pi | B \Phi_0 \right\rangle_{\rm flavor} = - \sqrt{\frac{3}{10}}  
	\end{array}  \mbox{ }.
\end{equation}

\end{appendix}

%%%%%%%%%%%%%%%%%%%%%%%%%%%%%%%%%%%%%%%%%%%%%%%%%%


\begin{thebibliography}{55}

\bibitem{Choi:2003ue}
  S.~K.~Choi {\it et al.}  [Belle Collaboration],
  %``Observation of a narrow charmonium - like state in exclusive B+- ---> K+- pi+ pi- J / psi decays,''
  Phys.\ Rev.\ Lett.\  {\bf 91}, 262001 (2003).
  
\bibitem{Aaij:2013zoa}
  R.~Aaij {\it et al.}  [LHCb Collaboration],
  %``Determination of the X(3872) meson quantum numbers,''
  Phys.\ Rev.\ Lett.\  {\bf 110}, 222001 (2013).  
  
\bibitem{Aad:2011ih} 
  G.~Aad {\it et al.}  [ATLAS Collaboration],
  %``Observation of a new $\chi_b$ state in radiative transitions to $\Upsilon(1S)$ and $\Upsilon(2S)$ at ATLAS,''
  Phys.\ Rev.\ Lett.\  {\bf 108}, 152001 (2012).  
  
\bibitem{Abazov:2012gh} 
  V.~M.~Abazov {\it et al.}  [D0 Collaboration],
  %``Observation of a narrow mass state decaying into $\Upsilon(1S) + \gamma$ in $p\bar{p}$ collisions at $\sqrt{s} = 1.96$ TeV,''
  Phys.\ Rev.\ D {\bf 86}, 031103 (2012).          
  
\bibitem{Aubert:2003fg} 
  B.~Aubert {\it et al.}  [BaBar Collaboration],
  %``Observation of a narrow meson decaying to $D_s^+ \pi^0$ at a mass of 2.32-GeV/c$^2$,''
  Phys.\ Rev.\ Lett.\  {\bf 90}, 242001 (2003).
  
\bibitem{Besson:2003cp} 
  D.~Besson {\it et al.}  [CLEO Collaboration],
  %``Observation of a narrow resonance of mass 2.46-GeV/c**2 decaying to D*+(s) pi0 and confirmation of the D*(sJ)(2317) state,''
  Phys.\ Rev.\ D {\bf 68}, 032002 (2003)
  [Phys.\ Rev.\ D {\bf 75}, 119908 (2007)].    
  
\bibitem{Bebek:1980zd} 
  C.~Bebek {\it et al.}  [CLEO Collaboration],
  %``Evidence for New Flavor Production at the Upsilon (4S),''
  Phys.\ Rev.\ Lett.\  {\bf 46}, 84 (1981).
  
\bibitem{LeeFranzini:1990gy} 
  J.~Lee-Franzini {\it et al.},
  %``Hyperfine splitting of B mesons and B(s) production at the Upsilon (5S),''
  Phys.\ Rev.\ Lett.\  {\bf 65}, 2947 (1990).
  
\bibitem{Abe:1998fb} 
  F.~Abe {\it et al.}  [CDF Collaboration],
  %``Observation of $B_c$ mesons in $p\bar{p}$ collisions at $\sqrt{s} = 1.8$ TeV,''
  Phys.\ Rev.\ D {\bf 58}, 112004 (1998).      

\bibitem{Nakamura:2010zzi}
  C. Patrignani {\it et al.} [Particle Data Group], 
  Chin. Phys. C {\bf 40}, 100001 (2016).
  
\bibitem{Brambilla:2010cs} 
  N.~Brambilla {\it et al.},
  %``Heavy quarkonium: progress, puzzles, and opportunities,''
  Eur.\ Phys.\ J.\ C {\bf 71}, 1534 (2011)
  
\bibitem{Link:2003bd} 
  J.~M.~Link {\it et al.}  [FOCUS Collaboration],
  %``Measurement of masses and widths of excited charm mesons D(2)* and evidence for broad states,''
  Phys.\ Lett.\ B {\bf 586}, 11 (2004).  
  
\bibitem{Abe:2003zm} 
  K.~Abe {\it et al.}  [Belle Collaboration],
  %``Study of B- ---> D**0 pi- (D**0 ---> D(*)+ pi-) decays,''
  Phys.\ Rev.\ D {\bf 69}, 112002 (2004).
  
\bibitem{Abazov:2007vq} 
  V.~M.~Abazov {\it et al.}  [D0 Collaboration],
  %``Observation and Properties of $L = 1 B_{1}$ and $B^*_2$ Mesons,''
  Phys.\ Rev.\ Lett.\  {\bf 99}, 172001 (2007).
  
\bibitem{Aaltonen:2008aa} 
  T.~Aaltonen {\it et al.}  [CDF Collaboration],
  %``Measurement of Resonance Parameters of Orbitally Excited Narrow $B^0$ Mesons,''
  Phys.\ Rev.\ Lett.\  {\bf 102}, 102003 (2009).    
  
\bibitem{delAmoSanchez:2010vq} 
  P.~del Amo Sanchez {\it et al.} [BaBar Collaboration],
  %``Observation of new resonances decaying to $D\pi$ and $D^*\pi$ in inclusive $e^+e^-$ collisions near $\sqrt{s}=$10.58 GeV,''
  Phys.\ Rev.\ D {\bf 82}, 111101 (2010).  
  
\bibitem{Aaij:2013sza} 
  R.~Aaij {\it et al.} [LHCb Collaboration],
  %``Study of $D_J$ meson decays to $D^+\pi^-$, $D^0 \pi^+$ and $D^{*+}\pi^-$ final states in pp collision,''
  JHEP {\bf 1309}, 145 (2013).  
  
\bibitem{Aaltonen:2013atp} 
  T.~A.~Aaltonen {\it et al.} [CDF Collaboration],
  %``Study of orbitally excited $B$ mesons and evidence for a new $B\pi$ resonance,''
  Phys.\ Rev.\ D {\bf 90}, no. 1, 012013 (2014).  
  
\bibitem{Kilian:1992hq} 
  U.~Kilian, J.~G.~Korner and D.~Pirjol,
  %``Excited charmed mesons in chiral perturbation theory,''
  Phys.\ Lett.\ B {\bf 288}, 360 (1992).  
  
\bibitem{Colangelo:1995ph} 
  P.~Colangelo, F.~De Fazio, G.~Nardulli, N.~Di Bartolomeo and R.~Gatto,
  %``Strong coupling of excited heavy mesons,''
  Phys.\ Rev.\ D {\bf 52}, 6422 (1995).
  
\bibitem{Casalbuoni:1996pg} 
  R.~Casalbuoni, A.~Deandrea, N.~Di Bartolomeo, R.~Gatto, F.~Feruglio and G.~Nardulli,
  %``Phenomenology of heavy meson chiral Lagrangians,''
  Phys.\ Rept.\  {\bf 281}, 145 (1997).    
  
\bibitem{Colangelo:2007ds} 
  P.~Colangelo, F.~De Fazio, S.~Nicotri and M.~Rizzi,
  %``Identifying D(sJ)(2700) through its decay modes,''
  Phys.\ Rev.\ D {\bf 77}, 014012 (2008).  
  
\bibitem{Wang:2010ydc}   
  Z.~G.~Wang,
  %``Analysis of strong decays of the charmed mesons D(2550), D(2600), D(2750) and D(2760),''
  Phys.\ Rev.\ D {\bf 83}, 014009 (2011).  
  
\bibitem{Colangelo:2012xi} 
  P.~Colangelo, F.~De Fazio, F.~Giannuzzi and S.~Nicotri,
  %``New meson spectroscopy with open charm and beauty,''
  Phys.\ Rev.\ D {\bf 86}, 054024 (2012).  
  
\bibitem{Wang:2013tka} 
  Z.~G.~Wang,
  %``Analysis of strong decays of the charmed mesons $D_J(2580), D^*_J(2650), D_J(2740), D^*_J(2760), D_J(3000), D^*_J(3000)$,''
  Phys.\ Rev.\ D {\bf 88}, no. 11, 114003 (2013).
  
\bibitem{Wang:2014cta} 
  Z.~G.~Wang,
  %``Strong decays of the bottom mesons $B_1(5721)$, $B_2(5747)$, $B_{s1}(5830)$, $B_{s2}(5840)$ and $B(5970)$,''
  Eur.\ Phys.\ J.\ Plus {\bf 129}, 186 (2014).
  
\bibitem{Wang:2016ewb} 
  Z.~G.~Wang,
  %``Strong decays of the charmed mesons $D_1^*(2680)$, $D^*_3(2760)$, $D_2^*(3000)$,''
  Commun.\ Theor.\ Phys.\  {\bf 66}, 671 (2016).
  
\bibitem{Capstick:2000qj} 
  S.~Capstick and W.~Roberts,
  %``Quark models of baryon masses and decays,''
  Prog.\ Part.\ Nucl.\ Phys.\  {\bf 45}, S241 (2000).     
    
\bibitem{Micu}
  L.~Micu,
  %``Decay rates of meson resonances in a quark model,''
  Nucl.\ Phys.\ B {\bf 10}, 521 (1969).
  
\bibitem{Kokoski:1985is}
  R.~Kokoski and N.~Isgur,
  %``Meson Decays by Flux Tube Breaking,''
  Phys.\ Rev.\ D {\bf 35}, 907 (1987).  
  
\bibitem{Kumano:1988ga}
  S.~Kumano and V.~R.~Pandharipande,
  %``Decay Of Mesons In Flux Tube Quark Model,''
  Phys.\ Rev.\ D {\bf 38}, 146 (1988).  
  
\bibitem{Stancu:1988gb}
  F.~Stancu and P.~Stassart,
  %``Pion Decay Of Baryons In A Flux Tube Quark Model,''
  Phys.\ Rev.\ D {\bf 38}, 233 (1988);
  %``Role Of The Pion Size And Flux Tube Extension In A Baryon Decay Model,''
  {\bf 39}, 343 (1989);
  %``Improved description of the Roper resonance in a constituent quark model,''
  {\bf 41}, 916 (1990);  
  %``N + rho decay of baryons in a flux tube breaking mechanism,''
  {\bf 42}, 1521 (1990).    
  
\bibitem{Geiger:1994}  
  P.~Geiger and E.~S.~Swanson, 
  Phys.\ Rev.\ D {\bf 50}, 6855 (1994).      
  
\bibitem{Kalashnikova:2005ui} 
  Y.~S.~Kalashnikova,
  %``Coupled-channel model for charmonium levels and an option for X(3872),''
  Phys.\ Rev.\ D {\bf 72}, 034010 (2005).  
  
\bibitem{Segovia:2012cd}
  J.~Segovia, D.~R.~Entem and F.~Fernandez,
  %``Scaling of the 3P0 Strength in Heavy Meson Strong Decays,''
  Phys.\ Lett.\ B {\bf 715}, 322 (2012).   
  
\bibitem{Blundell:1995ev}
  H.~G.~Blundell and S.~Godfrey,
  %``The Xi (2220) revisited: Strong decays of the 1(3) F2 1(3) F4 s anti-s mesons,''
  Phys.\ Rev.\ D {\bf 53}, 3700 (1996).  
  
\bibitem{Ackleh:1996yt} 
  E.~S.~Ackleh, T.~Barnes and E.~S.~Swanson,
  %``On the mechanism of open flavor strong decays,''
  Phys.\ Rev.\ D {\bf 54}, 6811 (1996);
  T.~Barnes, F.~E.~Close, P.~R.~Page and E.~S.~Swanson,
  %``Higher quarkonia,''
  Phys.\ Rev.\ D {\bf 55}, 4157 (1997).  
  
\bibitem{LeYaouanc}  
  A.~Le Yaouanc, L.~Oliver, O.~Pene and J.~-C.~Raynal,
  %``Naive quark pair creation model of strong interaction vertices,''
  Phys.\ Rev.\ D {\bf 8}, 2223 (1973);
  %``Naive quark pair creation model and baryon decays,''
  {\bf 9}, 1415 (1974).    
  
\bibitem{Capstick:1992th} 
  S.~Capstick and W.~Roberts,
  %``N pi decays of baryons in a relativized model,''
  Phys.\ Rev.\ D {\bf 47}, 1994 (1993);
  %``Quasi two-body decays of nonstrange baryons,''
  {\bf 49}, 4570 (1994);
  %``Strange decays of nonstrange baryons,''
  {\bf 58}, 074011 (1998).  
  
 \bibitem{Strong2015}  
  R.~Bijker, J.~Ferretti, G. Galat\`a, H. ~Garc\'ia-Tecocoatzi and E.~Santopinto,
  Phys. Rev. D {\bf 94}, 074040 (2016);
  H.~Garc\'ia-Tecocoatzi, R.~Bijker, J.~Ferretti, G.~Galat\`a and E.~Santopinto,
  %``Open flavor strong decays,''
  Few-Body Syst. {\bf 57}, 985 (2016).  
  
\bibitem{LeYaouanc02}  
  A.~Le Yaouanc, L.~Oliver, O.~Pene and J.~-C.~Raynal,
  %``Strong Decays of psi-prime-prime (4.028) as a Radial Excitation of Charmonium,''
  Phys.\ Lett.\ B {\bf 71}, 397 (1977);
  %``Why Is psi-prime-prime-prime (4.414) SO Narrow?,''
  {\bf 72}, 57 (1977).    
  
\bibitem{Barnes:2005pb}
  T.~Barnes, S.~Godfrey and E.~S.~Swanson,
  %``Higher charmonia,''
  Phys.\ Rev.\ D {\bf 72}, 054026 (2005).  
  
\bibitem{Ferretti:2013faa}   
   J.~Ferretti, G.~Galat\`a and E.~Santopinto,
  %``Interpretation of the X(3872) as a charmonium state plus an extra component due to the coupling to the meson-meson continuum,''
  Phys.\ Rev.\ C {\bf 88}, 015207 (2013). 
	
\bibitem{Ferretti:2014xqa}   
  J.~Ferretti, G.~Galat\`a and E.~Santopinto,
  %``Quark structure of the $X(3872)$ and $\chi_b(3P)$ resonances,''
  Phys.\ Rev.\ D {\bf 90}, 054010 (2014).  
  
\bibitem{Ferretti:2013vua}    
  J.~Ferretti and E.~Santopinto,
  %``Higher mass bottomonia,''
  Phys.\ Rev.\ D {\bf 90}, 094022 (2014).  
  
\bibitem{Godfrey:2015dia} 
  S.~Godfrey and K.~Moats,
  %``Bottomonium Mesons and Strategies for their Observation,''
  Phys.\ Rev.\ D {\bf 92}, no. 5, 054034 (2015).  
  
\bibitem{Close:2005se}
  F.~E.~Close and E.~S.~Swanson,
  %``Dynamics and decay of heavy-light hadrons,''
  Phys.\ Rev.\ D {\bf 72}, 094004 (2005).   
  
\bibitem{Godfrey:2015dva} 
  S.~Godfrey and K.~Moats,
  %``Properties of Excited Charm and Charm-Strange Mesons,''
  Phys.\ Rev.\ D {\bf 93}, no. 3, 034035 (2016). 
  
\bibitem{Godfrey:2016nwn} 
  S.~Godfrey, K.~Moats and E.~S.~Swanson,
  %``$B$ and $B_s$ Meson Spectroscopy,''
  Phys.\ Rev.\ D {\bf 94}, no. 5, 054025 (2016).  
  
 \bibitem{Godfrey:1985xj}  
  S.~Godfrey and N.~Isgur,
  %``Mesons in a Relativized Quark Model with Chromodynamics,''
  Phys.\ Rev.\ D {\bf 32}, 189 (1985).

\bibitem{Godfrey:2004ya} 
  S.~Godfrey,
  %``Spectroscopy of $B_c$ mesons in the relativized quark model,''
  Phys.\ Rev.\ D {\bf 70}, 054017 (2004).   
  
\bibitem{bottomonium}
  J.~Ferretti, G.~Galat\'a, E.~Santopinto and A.~Vassallo,
  Phys.\ Rev.\ C {\bf 86}, 015204 (2012).
  
\bibitem{Hwang:2004cd} 
  D.~S.~Hwang and D.~W.~Kim,
  %``Mass of D*(sJ)(2317) and coupled channel effect,''
  Phys.\ Lett.\ B {\bf 601}, 137 (2004). 
  
\bibitem{Guo:2011dv} 
  F.~K.~Guo and U.~G.~Mei{\ss}ner,
  %``Examining coupled-channel effects in radiative charmonium transitions,''
  Phys.\ Rev.\ Lett.\  {\bf 108}, 112002 (2012).
  
\bibitem{Cardoso:2014xda} 
  M.~Cardoso, G.~Rupp and E.~van Beveren,
  %``Unquenched quark-model calculation of $X(3872)$ electromagnetic decays,''
  Eur.\ Phys.\ J.\ C {\bf 75}, no. 1, 26 (2015).  
    
\bibitem{Guo:2016yxl} 
  F.~K.~Guo, U.~G.~Mei{\ss}ner and Z.~Yang,
  %``Hindered magnetic dipole transitions between P-wave bottomonia and coupled-channel effects,''
  Phys.\ Lett.\ B {\bf 760}, 417 (2016).    
  
\bibitem{Lu:2016mbb} 
  Y.~Lu, M.~N.~Anwar and B.~S.~Zou,
  %``Coupled-Channel Effects for the Bottomonium with Realistic Wave Functions,''
  Phys.\ Rev.\ D {\bf 94}, no. 3, 034021 (2016).  
  
\bibitem{Ferretti:2018tco} 
  J.~Ferretti and E.~Santopinto,
  %``Threshold effects in $\chi_{\rm c}(2P)$ and $\chi_{\rm b}(3P)$ multiplets and $J/\Psi \rho, J/\Psi \omega$ hidden-flavor strong decays of the $X(3872)$,''
  arXiv:1806.02489.  
  
\bibitem{Roberts:1992}
  W.~Roberts and B.~Silvestre-Brac,
  Few-Body Syst. {\bf 11}, 171 (1992).  
  
\bibitem{self-energies}
  H. ~Garc\'ia-Tecocoatzi, R.~Bijker, J.~Ferretti and E.~Santopinto,
  Eur. J. Phys. A {\bf 53}, 115 (2017).  
  
\bibitem{Albrecht:1989pa} 
  H.~Albrecht {\it et al.} [ARGUS Collaboration],
  %``Resonance Decomposition of the $D^*$0 (2420) Through a Decay Angular Analysis,''
  Phys.\ Lett.\ B {\bf 232}, 398 (1989).  
  
\bibitem{Avery:1994yc} 
  P.~Avery {\it et al.} [CLEO Collaboration],
  %``Production and decay of D$_1$(2420)$^0$ and D$_2^*$(2460)$^0$,''
  Phys.\ Lett.\ B {\bf 331}, 236 (1994).  
  
\bibitem{Chekanov:2008ac} 
  S.~Chekanov {\it et al.} [ZEUS Collaboration],
  %``Production of excited charm and charm-strange mesons at HERA,''
  Eur.\ Phys.\ J.\ C {\bf 60}, 25 (2009).  
  
\bibitem{Aubert:2009ah} 
  B.~Aubert {\it et al.} [BaBar Collaboration],
  %``Study of D(sJ) decays to D*K in inclusive e+ e- interactions,''
  Phys.\ Rev.\ D {\bf 80}, 092003 (2009).  
  
\bibitem{Aaij:2015qla} 
  R.~Aaij {\it et al.} [LHCb Collaboration],
  %``Precise measurements of the properties of the $B_1(5721)^{0,+}$ and $B^\ast_2(5747)^{0,+}$ states and observation of $B^{+,0}\pi^{-,+}$ mass structures,''
  JHEP {\bf 1504}, 024 (2015).  
  
\bibitem{Geiger-Isgur} 
  P.~Geiger and N.~Isgur,
  %``How the Okubo-Zweig-Iizuka rule evades large loop corrections,''
  Phys.\ Rev.\ Lett.\  {\bf 67}, 1066 (1991);
   %``Reconciling the OZI rule with strong pair creation,''
  Phys.\ Rev.\  D {\bf 44}, 799 (1991);
  %``When can hadronic loops scuttle the OZI rule?,''
  {\bf 47}, 5050 (1993).  
  
\bibitem{Geiger:1996re}
  P.~Geiger and N.~Isgur,
  %``Strange hadronic loops of the proton: A quark model calculation,''
  Phys.\ Rev.\  D {\bf 55}, 299 (1997).

\bibitem{Bijker:2009up}
  R.~Bijker and E.~Santopinto,
  %``Unquenched quark model for baryons: Magnetic moments, spins and orbital angular momenta,''
  Phys.\ Rev.\ C {\bf 80}, 065210 (2009).

\bibitem{Santopinto:2010zza}   
  E.~Santopinto and R.~Bijker,
  %``Flavor asymmetry of sea quarks in the unquenched quark model,''
  Phys.\ Rev.\ C {\bf 82}, 062202 (2010).
  
\bibitem{Bijker:2012zza}    
  R.~Bijker, J.~Ferretti and E.~Santopinto,
  %``$s\bar{s}$ sea pair contribution to electromagnetic observables of the proton in the unquenched quark model,''
  Phys.\ Rev.\ C {\bf 85}, 035204 (2012).                                 
			
\end{thebibliography}
\end{document}